\documentclass[sigplan,screen]{acmart}
\usepackage{natbib}

\usepackage{amssymb}
\newcommand{\NeedsSchema}{\textsf{NeedsSchema}}

\usepackage{listings}
\usepackage{xcolor}
\newcommand{\RefinedT}[2]{\{\,#1 \mid #2\,\}}
\newcommand{\Satisfied}{\mathsf{sat}}

\begin{document}

\title[Extended Abstract]{Type-Driven Prompt Programming: From Typed Interfaces to a Calculus of Constraints} 
\author{Abhijit Paul}
\authornotemark[1]
\email{bsse1201@iit.du.ac.bd}
\affiliation{%
  \institution{University of Dhaka}
  \country{Bangladesh}
}

\renewcommand{\shortauthors}{Paul et al.}

\begin{abstract}
Prompt programming treats LLM prompts as software components with typed interfaces. Through a literature survey of 15 recent works (2023-2025), we observe a consistent trend: type systems are central to emerging prompt programming frameworks. However, there are gaps in constraint expressiveness and algorithms. To address it, we introduce the notion of $\lambda$ Prompt, a dependently typed calculus with probabilistic refinements for syntactic/semantic constraints. While not yet a full calculus, our formulation motivates a type-theoretic foundation for prompt programming. Our catalog of 13 constraints reveals underexplored areas in constraint expressiveness (C9–C13). To address algorithmic gap, we propose a constraint-preserving optimization rule. Finally, we outline research directions on prompt program compiler.
\end{abstract}
\maketitle

\section{Introduction}
Prompt engineering is in high demand, with a recent industry report estimating that over 100 million prompts are generated globally each day \cite{byteplus2024prompts}. As large language models (LLM) are being integrated into software systems, their prompts are becoming an essential component of these software systems. As a result, these prompts have the same robustness, security and type safety requirements that any software component has \cite{guy2024prompts}. So in this extended abstract, we focus on exploring an emerging philosophy for prompt engineering called "Prompt Programming" \cite{guy2024prompts}. This philosophy treats prompts as a software component and thus, works towards enforcing these robustness, security and type-safety requirements. Methodologically, we conducted a literature survey on prompt programs to explore and answer the following questions. 
\begin{itemize}
    \item RQ1. How is \textit{prompt programming} formally defined and characterized in current research literature?
    \item RQ2. What constitutes \textit{type-driven prompt programming}, and what factors explain its prevalence in contemporary prompt program literature?
    \item RQ3. What significant research gaps persist in type-driven approaches to prompt programming?
\end{itemize}

\section{Literature Survey} \label{section:literature-survey}
Prompt programming is an emerging domain so we were able to find a total of 7 prompt-programming frameworks and 8 empirical studies published between 2023-2025. Papers were selected via keyword searches (“prompt engineering”, “typed prompts”, “prompt engineering”) on arXiv, ACL Anthology, and major GitHub repos, then filtered for those proposing structured, reusable interfaces. Given the concise nature of extended abstracts, we forgo detailed examination of individual works and instead focus on high-level view. 

An interesting observation emerged from our literature survey: all of the identified papers and frameworks on prompt programs \cite{baml2024,typechat2023, llmexe2024, instructor2024, fructose2024, beurer2023prompting, liang2024prompts, guy2024prompts} employ typed prompts. The sole exception, the work of Tobias and Neville et al., utilizes a symbolic representation of structured prompts \cite{schnabel2024prompts, schnabel2024symbolic2}, which can be compared to a well-typed structure. This finding led us to explore the role of type system in prompt programming. It appeared that whenever researchers adopted a structured, pragmatic approach to prompt programming, they inevitably incorporated a type system. This consistent trend suggests that type systems may play a crucial role in the development of prompt programming, a dimension that has yet to be fully explored. This realization forms the foundation of our research and motivated the writing of this paper.

\section{Prompt Programming (RQ1)} \label{section:RQ1} 
Prompt programming began from the notion that prompts themselves can be treated as programs \cite{guy2024prompts}. We examine how prompt programming is perceived by both the academia and industry. Academia has a broad view of prompt programs. According to Liang, Jenny et al, prompt programs are prompts that can accept variable inputs and could be interpreted by an LLM to perform specified actions and/or generate output. This prompt is executed within a software application or code by a LLM \cite{liang2024prompts}. Observe that, this definition excludes single-use prompts where an user converses with an LLM to achieve a goal e.g. debugging. Beurer-Kellner, Fischer et al \cite{beurer2023prompting} defines a structured language for prompt programming, while Tobias and Neville introduces the notion of program optimization into prompt programs \cite{schnabel2024prompts, schnabel2024symbolic2}. On the other hand, industry and open-source community tends to universally treat prompt programs as well-typed functions \cite{baml2024, beurer2023prompting, llmexe2024, instructor2024, fructose2024}. Tools like Typechat \cite{typechat2023} and DSPy \cite{dspy2024} goes beyond simple type validation and guide LLMs to produce correctly typed outputs in case of errors.

Drawing from these views, we propose the $\lambda$Prompt calculus—a dependently typed calculus specifically tailored to model prompt programs with probabilistic refinements. 

Formally, a prompt program is a 4-tuple $(I, O, P, C)$:

\begin{itemize}
\item $I, O$ : Dependent types $\Sigma x : \tau. \varphi(x)$, where $\tau$ is the base type (e.g., String, JSON) and $x$ ranges over its values, refined by a predicate $\varphi$ that may be probabilistic or semantic.
\item $P$: Natural-language instructions as effectful computations via interaction with LLM, denoted as $LLM \ \varepsilon (I \to O)$, capturing the stochastic nature of LLM responses.
\item $C$ : Constraints $c_1, \dots, c_n$ from Table \ref{tab:llm_constraints}, as refinements.
\end{itemize}

\begin{table}[ht]
\centering
\caption{Constraints for Prompt Programs.}
\label{tab:llm_constraints}
\begin{tabular}{p{3cm} p{5cm}}
\toprule
\textbf{Constraint Type} & \textbf{Description and Example} \\
\midrule
C1. Domain-Specific Constraints \cite{yang2024plug} & Output sequence must follow domain syntax e.g. LTL in Robotics \\

C2. Structured Output Constraint \cite{liu2024we} & Enforce structures like markdown, HTML, DSL. \\

C3. Decoding Constraints \cite{schnabel2024prompts} & Modify generation process (e.g., constrained decoding) to satisfy output forms or topics. \\

C4. JSON Schema Constraint \cite{liu2024we} & Output must match a predefined JSON structure \\

C5. Label Range Constraint  \cite{liu2024we} &  Constrain answers to fixed options. \emph{E.g.,} Positive/Negative/Neutral. \\

C6. Length Constraints \cite{liu2024we} & Restrict number of tokens. \emph{E.g.,} Bullet < 40 words. \\

C7. Exclusion Constraints \cite{liu2024we} & Forbid certain content. \emph{E.g.,} Remove PII and boilerplate HTML. \\

C8. Inclusion Constraints \cite{liu2024we} & Require certain content. \emph{E.g.,} Email must mention manager and office. \\

C9. Domain Constraints \cite{liu2024we} & Limit to specific subject areas. \emph{E.g.,} Discuss Airtel but not its competitors. \\

C10. Tone Constraints \cite{liu2024we} & Require specific tone \emph{E.g.,} Friendly voice, layman’s terms. \\

C11. Input Sanitation \cite{chong2024casper} & Santize input data. Often trivial. \\

C12. Encoding Constraints \cite{wang2022integrating} & Apply lexical, Ontogical constraints on input encoding. \\


C13. Mental Model Constraint \cite{blokpoel_vanrooij_2021} &  LLM's behavior conforming to developer's mental model by constraining it\\

\bottomrule
\end{tabular}
\vspace{-5pt}
\end{table}
 
\begin{figure}
    \centering
    \includegraphics[width=0.9\linewidth]{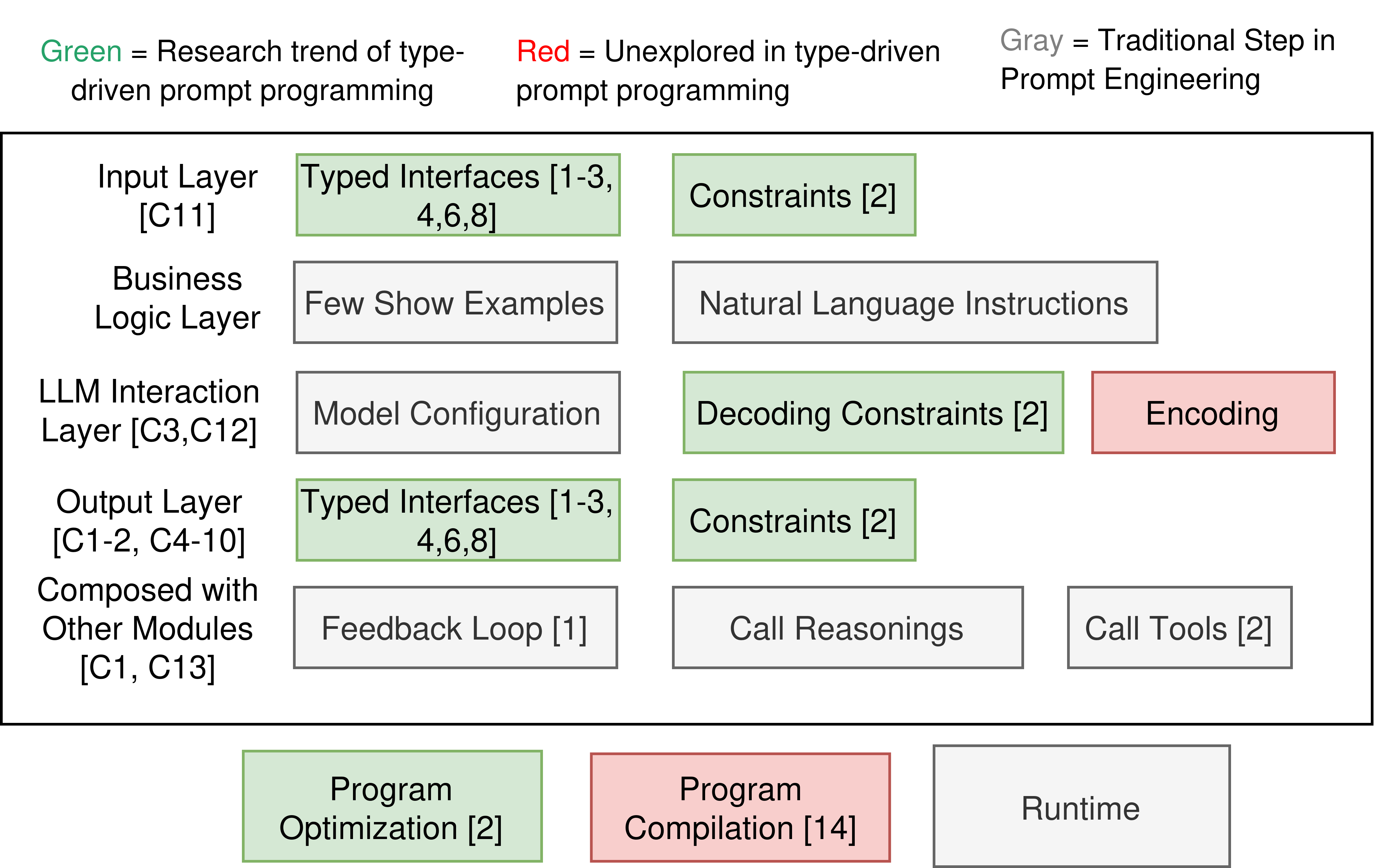}
    \caption{Layered Structure of a Prompt Program. (The C1-C13 codes refer to the constraints identified in Table-\ref{tab:llm_constraints})}
    \label{fig:enter-label}
\end{figure}

\section{Type-driven Prompt Programming (RQ2)}
As discussed in Section-\ref{section:literature-survey}, we had a surprising finding - we observed that all identified papers and frameworks on prompt programs directly \cite{baml2024,typechat2023, llmexe2024, instructor2024, fructose2024,beurer2023prompting, liang2024prompts, guy2024prompts} or indirectly \cite{schnabel2024prompts, schnabel2024symbolic2} use typed prompts. Such a research trend led us to the following questions: Why is this trend occurring? What motivates researchers and practitioners to employ type system in prompt programming? To answer these questions, we explored literature to identify major pain points in prompt engineering and the usage of type system as a solution to these pain points.

\subsection{Input and Output Types of Data} \label{subsection:common-issue}
Prompt programs are inside a large software. They take specific types of data as input and expects specific types of data (e.g. json, xml etc) as output from the LLM. Very often, LLM generates data that does not conform to the type required by the software \cite{ugare2024syncode}. So ensuring that output data conforms to the defined type is a critical need. To address it, developers have proposed libraries such as TypeChat, llm-exe, Instructor(pydantic), Fructose to define and validate types of output data \cite{typechat2023, llmexe2024, instructor2024, fructose2024}. 
\subsection{Maintainability of Codebase}
Prompts written using long f-strings with dynamic inserts are like inline scripts - there's no separation of logic and structure. It becomes difficult to maintain across versions and files \cite{baml2024}. This has led practitioners to express prompts as typed functions, where functional decomposition resolves the issue of long f-strings, and typed interfaces address the concerns outlined in Section-\ref{subsection:common-issue}.

\subsection{Prompt Program Optimization}
Optimizing prompts is a major pain point. To address it, DsPy optimizes reasoning techniques and augments prompts on the fly \cite{dspy2024}. Tobias and Neville \cite{schnabel2024prompts, schnabel2024symbolic2} introduce a grid-search approach over a prompt program translated into a Directed Acyclic Graph (DAG). This allows them to automatically apply a set of mutations to reach an optimal prompt.

\vspace{-1em} 
\subsection{Other Pain Points}
Liang, Jenny et al has identified 4 major paint points. Firstly, each LLM has its own limitations. This adds complexity to prompt engineering. Secondly, LLMs have stochastic behavior. Engineering prompts by interacting with such a stochastic system is tedious. Thirdly, minute details in the prompt matter. Prompts are finicky and fragile. This adds uncertainty in prompt engineering. And finally, testing prompt behaviors is costly so testing is often limited or time-consuming \cite{liang2024prompts}. We note that the libraries and research papers identified in our literature survey does not solve either of these issues. To the best of our knowledge, this remains an open research direction for type-driven prompt programming.

\section{Research Gaps (RQ3)}\label{section:research-gap} 
To systematically identify research gaps, we first stratify the current literature into a prompt-program structure, identified in Section-\ref{section:RQ1}.
As we can see in Figure-\ref{fig:enter-label}, there has been no work on incorporating encoding-level constraints for prompt programs. Additionally, there has been few works on prompt program optimization \cite{schnabel2024prompts, schnabel2024symbolic2}. Specifically, the authors used structure-aware optimization. They however did not consider constraints over types. Additionally, constraints used in \cite{beurer2023prompting} are syntactic in nature and not semantic. Similarly, typed interfaces in input and output layers in Figure-\ref{fig:enter-label} are also syntactic in nature \cite{baml2024,typechat2023, llmexe2024, instructor2024, fructose2024, beurer2023prompting, liang2024prompts, guy2024prompts}. This showcases two crucial gaps - limited expressiveness of constraints and limited use of algorithms in prompt programs.
\subsection{Constraints Expressiveness}
The diversity of prompt program constraints addressed in literature is limited. To address this gap, we explore literature on user needs for input-output validation \cite{liu2024we}, prompt engineering challenges \cite{liang2024prompts} and emerging doamins \cite{blokpoel_vanrooij_2021} to identify the following 13 constraints for prompt programs in Table-\ref{tab:llm_constraints}. Note that users have expressed their needs for constraints C9, C10, C13 in \cite{liu2024we, blokpoel_vanrooij_2021} but these constraints remain mostly unexplored. We note that these semantic constraints (C9, C10, C13) map to type refinements:

\begin{itemize}
    \item C10 (Tone): $\{s : \text{String} \mid \text{Formality}(s) \geq 0.7\}$
    \item C13 (Mental Model): $\{f : I \to O \mid \forall x. P_\delta(f(x) \approx \text{human\_expectation}(x))\}$
    \item C9 (Domain): Ontology-parametrized type: \\ 
    $\text{PromptType(Finance)} \to \text{JSON}$
\end{itemize}
These functions can be small language models (SLM). 
With recent works in speculative decoding \cite{leviathan2023fast}, such SLMs won't impose much performance penalty. Other constraints are detailed in prior work and are omitted here for brevity.

\subsection{Algorithmic Limitations} 
Prompt programs are programs, so they require compilation, optimization and runtime handling algorithm. DsPy proposes a compiler that expresses prompting, finetuning, augmentation, and reasoning techniques as parameterized modules. Thus it can learn to apply the best composition of techniques for executing a prompt \cite{khattab2023dspy}. Tobias and Neville et al works on prompt program optimization \cite{schnabel2024prompts}. But it does not impose the type constraints during optimization. So we extend Tobias et al.'s DAG optimization \cite{schnabel2024symbolic2} with \emph{constraint-aware rewriting}. For a prompt program $e$ typed
$(\tau, c)$  optimization is defined as
\[
  \text{optimize}(e) \;=\;
  \underset{e' \in \mathcal{M}(e)}{\arg\min}\;
  \mathbb{E}_{x\sim\mathcal{D}}
  \bigl[\,\text{cost}(e',x)\bigr]
  \quad\text{s.t.}\quad
  \Gamma \vdash e' : \tau
  \;\land\;
  \text{sat}(e',c).
\]
where $\mathcal{M}$ = structure-preserving mutations, $\text{sat}$ = semantic satisfiability of constraint $c$ and $\text{cost}$ = prediction error plus compute cost.  

\paragraph{Constraint-guided pruning.}
Let $c = \NeedsSchema$ (“prompt must embed an explicit JSON schema”).  
We specialize the mutation set via the rule
\[
  \frac{
        \Gamma \,\vdash\, e : \RefinedT(\tau,\NeedsSchema)
        \quad
        sch \sim \llbracket \tau \rrbracket_{\text{schema}}
      }{
        \mathcal{M}_{\text{type}}(e)
        \;=\;
        \{\, e \,\oplus_{k} sch
        \mid
        k \in \text{schema\_slots}(e)
        \}
        }
\]
Here $sch$ is a sampled, well-formed schema and
$\oplus_{k}$ injects it at slot $k$.
The search space drops from
$\mathcal{O}(|\mathcal{M}|)$
to
$\mathcal{O}(|\text{schema\_slots}|)$.

\begin{theorem}[Type preservation]
For every $e' \in \mathcal{M}_{\text{type}}(e)$,
\(
  \Gamma \,\vdash\, e' : \RefinedT(\tau,\Satisfied).
\)
\end{theorem}

Thus constraint tags steer the optimizer,pruning large portions of the search space. We'll experiment in future work.
\section*{Discussion}
This work was initiated by the surprising discovery of the growing trend on typed interfaces in prompt programming research. This observation led us to investigate why this trend is emerging and what motivates both researchers and practitioners to adopt type systems in prompt programming. To address these questions, we first provided a clear definition of prompt programming and then examined the common challenges users face when working with prompts. Type-driven prompt programming demands gradual verification: static types for syntactic constraints (C1–C8), probabilistic checks for semantic ones (C9–C13). Our notion of $\lambda$ Prompt calculus reduces prompt fragility by encoding the 13 constraints as refinements and guaranteeing optimization safety via type preservation. Moving forward, we aim to explore how semantic constraints can be effectively integrated into the type system, and eventually develop a compiler based on the structure of prompt programs.

\bibliographystyle{ACM-Reference-Format}
\bibliography{sample-base}
\end{document}